\documentstyle[aps,prl,preprint,floats,epsfig]{revtex}

\begin{document}

\preprint{\tighten\vbox{\hbox{\hfil CLNS 00/1679}
                        \hbox{\hfil CLEO 00-11}
}}

\title{
Search for Decays of $B^0$ Mesons Into Pairs of Leptons:
$B^0\rightarrow e^+e^-, B^0\rightarrow \mu^+\mu^-$ 
and $B^0\rightarrow e^{\pm}\mu^{\mp}$.
 }  

\author{CLEO Collaboration}
\date{\today}

\maketitle
\tighten

\begin{abstract} 
We search for the decay of the $B^0$ meson into a pair 
of leptons in the suppressed channels $B^0\rightarrow e^+e^-$,
$B^0\rightarrow \mu^+\mu^-$ and in the lepton number violating channel 
$B^0\rightarrow e^{\pm}\mu^{\mp}$ in a sample
of $9.7\times 10^6$ $B{\bar B}$ pairs
recorded by CLEO detector.
No signal is found, and the following upper limits on the 
branching fractions are established:
${\cal B}(B^0\rightarrow e^+e^-)< 8.3\times 10^{-7}$, 
${\cal B}(B^0\rightarrow  \mu^+\mu^-)< 6.1\times 10^{-7}$,
${\cal B}(B^0\rightarrow e^{\pm}\mu^{\mp})< 15\times 10^{-7}$ at 
90\% confidence level.
A new lower limit 
on the Pati-Salam leptoquark mass $M_{LQ}>27~\rm{TeV}$ is
established at 90\% confidence level.

\end{abstract}
\newpage

{
\renewcommand{\thefootnote}{\fnsymbol{footnote}}


\begin{center}
T.~Bergfeld,$^{1}$ B.~I.~Eisenstein,$^{1}$ J.~Ernst,$^{1}$
G.~E.~Gladding,$^{1}$ G.~D.~Gollin,$^{1}$ R.~M.~Hans,$^{1}$
E.~Johnson,$^{1}$ I.~Karliner,$^{1}$ M.~A.~Marsh,$^{1}$
M.~Palmer,$^{1}$
% ...
\end{center}
 
\small
\begin{center}
$^{1}${University of Illinois, Urbana-Champaign, Illinois 61801}\\
% ...
\end{center}

\setcounter{footnote}{0}
}
\newpage



\section{Introduction}

\begin{figure}
  \begin{center}
   \mbox{\epsfig{file=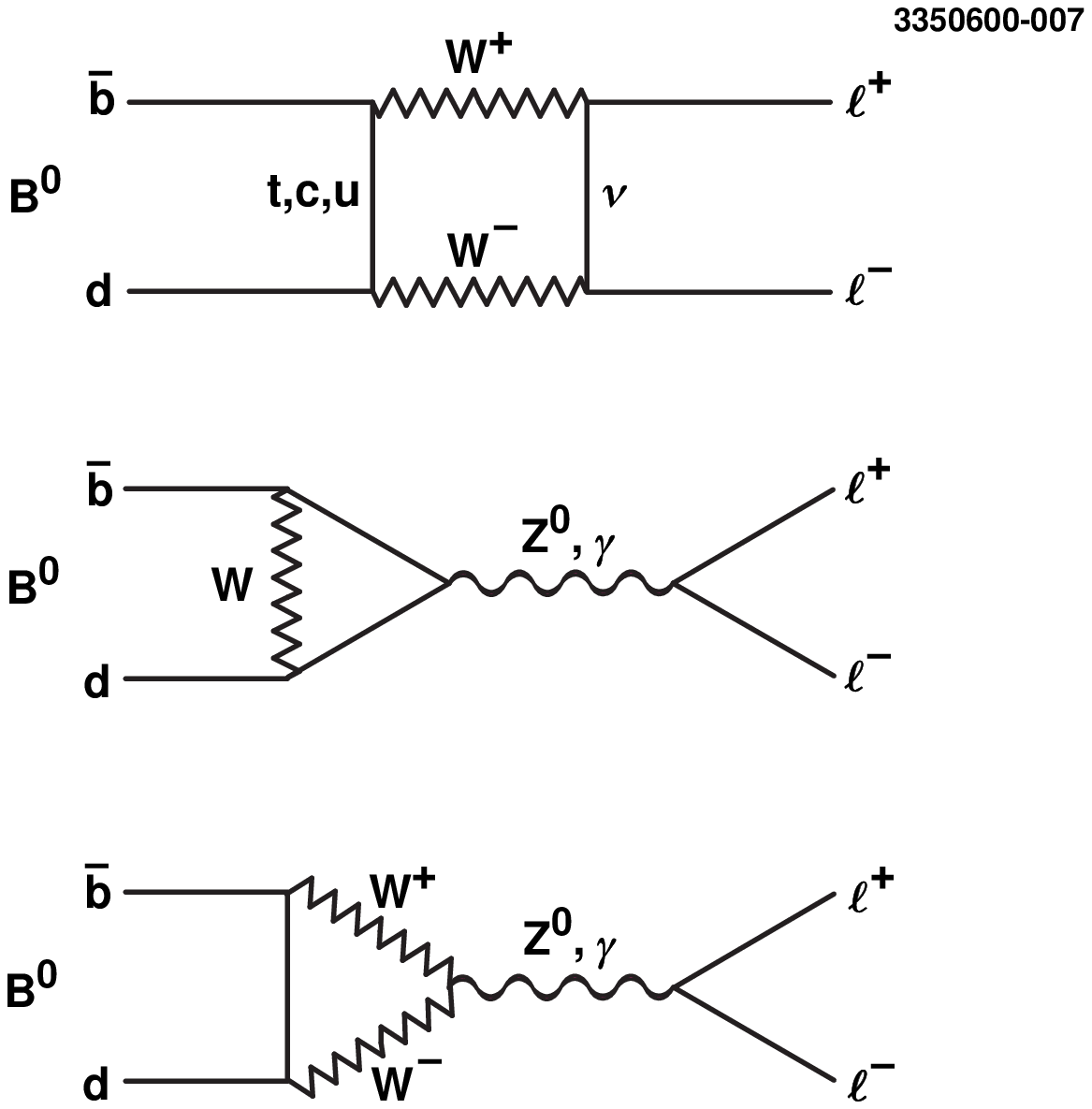,height=10cm}}
\caption{
\label{diagram} Feynman diagrams for the decay
$B^0\to l^+l^-$
($l=e,\mu$) in  the Standard Model.   }
  \end{center}
\end{figure}

In the Standard Model the decays $B^0\rightarrow e^+e^-$ and 
$B^0\rightarrow \mu^+\mu^-$ are allowed via the diagrams 
shown in Fig~\ref{diagram}.  
The Standard Model predictions for the branching fractions 
are ${\cal B}(B^0\rightarrow e^+e^-)=1.9\times 10^{-15}$ 
and ${\cal B}(B^0\rightarrow \mu^+\mu^-)=8.0\times 10^{-11}$, 
respectively\cite{SM-br}. These rates put them beyond the reach of 
current experiments. Their observation at CLEO would provide  
clear evidence for physics beyond the Standard Model.

In two Higgs doublet models, the corresponding branching
fractions can be strongly enhanced due to additional 
diagrams involving  Higgs bosons \cite{Rizzo}.
Supersymmetric particles can further modify the expectation, 
but calculations of SUSY contributions have not yet been made.
In some models with $Z$-mediated flavor changing neutral currents
\cite{Nir} the corresponding branching fraction can be enhanced 
by a factor of 400\cite{Gronau} compared to Standard Model predictions.

The decay  $B^0\rightarrow e^{\pm}\mu^{\mp}$ is forbidden in
the Standard Model by lepton number conservation. However it
can occur in Pati-Salam leptoquark models\cite{Pati-Salam}, due to 
the existence of particles that couple to both quarks and leptons.

Currently the  best 90\% confidence level 
limits on the  branching fractions 
are ${\cal B}(B^0\rightarrow e^+e^-)<59\times 10^{-7}$
(from CLEO\cite{lingel-paper}),
${\cal B}(B^0\rightarrow e^{\pm}\mu^{\mp})<35\times 10^{-7}$
(from CDF\cite{CDFem}) and
${\cal B}(B^0\rightarrow \mu^+\mu^-)<6.8\times 10^{-7}$ 
(from CDF\cite{CDFmm}).

\section{Data sample}

We use data taken with the CLEO II/II.V detector\cite{cleo_detector}
operating at the Cornell Electron Storage Ring (CESR). The sample 
consists of 9.1 $fb^{-1}$ taken on the $\Upsilon (4S)$ resonance,
corresponding to  approximately $9.7 \times  10^6~B{\bar B}$ pairs. 
An additional sample of 4.5 $fb^{-1}$ accumulated 
approximately 60 MeV below the 
$\Upsilon (4S)$ resonance is used to estimate the non-resonant 
background. Charge conjugation is implied throughout this paper
and equal production of charged and neutral $B$ mesons at
the $\Upsilon(4S)$ resonance is assumed.
The CLEO II charged particle tracking system consists of a 6-layer straw 
tube chamber, a 10-layer precision tracking chamber and a 51 layer main drift 
chamber. In the CLEO II.V configuration the straw tube chamber was replaced 
by a 3-layer, double-sided silicon tracker.  Beyond the tracking chambers, but 
inside the 1.5T solenoid magnet, are a time-of-flight system 
and an electromagnetic 
calorimeter consisting of 7800 CsI crystals. Electron 
identification is performed with a likelihood function combining
calorimeter ($E/p$, the ratio of energy deposited by particle 
in calorimeter to its momentum) 
and specific ionization ($dE/dx$) information. Muons are 
identified using proportional counters placed at various depths in the 
steel return yoke of the magnet.

Since, at a symmetric $e^+e^-$ collider, 
$B$ mesons at the $\Upsilon (4S)$ resonance are produced nearly at 
rest, 
the decays $B^0\rightarrow e^+e^-$, $B^0\rightarrow \mu^+\mu^-$ and
$B^0\rightarrow e^{\pm}\mu^{\mp}$ will contain two nearly
back-to-back leptons with momenta $p\approx 2.6~\rm{GeV}$.
In order to select $B{\bar B}$ events, we require at least five charged 
tracks in the event with two well reconstructed and oppositely charged tracks 
identified as leptons. 
The lepton pair is required to have a mass within three standard deviations
from the 
the mass of the $B^0$ meson. We use the beam constrained mass, 
$M(B)=
\sqrt{E_{beam}^2-{({\vec p_{l^+}}+{\vec p_{l^-}})}^2}$. The resolution on 
$M(B)$
is 2.8~MeV for the $B^0\rightarrow e^+e^-$ decay 
and 2.6~MeV for the $B^0\rightarrow \mu^+\mu^-$ and
$B^0\rightarrow e^{\pm}\mu^{\mp}$ decays. For signal events,  $M(B)$
is required to be within 3 standard deviations of the $B^0$ meson mass.

We calculate the energy difference between the candidate $B^0$ meson  
and the beam energy $\Delta E=E_{l^+}+E_{l^-}-E_{beam}$ and require 
that
the difference should be
$\vert\Delta E\vert < 75\mbox{ MeV}$
for $B^0\rightarrow \mu^+\mu^-$ decay, 
and for 
$B^0\rightarrow e^+e^-$ and $B^0\rightarrow e^{\pm}\mu^{\mp}$ decays 
it should satisfy $-100 \mbox{ MeV}< \Delta E < 75 \mbox{ MeV}$. 
The asymmetric $\Delta E$ requirement in  $B^0\rightarrow e^+e^-$ and 
$B^0\rightarrow e^{\pm}\mu^{\mp}$ channels takes into account  energy loss 
due to final state radiation from the electrons.

The main background comes from the processes 
$e^+e^-\to q{\bar q} (q=u,c,s,d)$ and 
$e^+e^-\to \tau^+\tau^-$.  Such events usually exhibit
a two jet structure and produce high momentum, approximately back-to-back
tracks which satisfy the requirements imposed on our candidate events.
We calculate the angle $\theta_{T}$ between 
the thrust axis of the
$B^0$ candidate tracks and the thrust axis of the remaining charged
and neutral tracks in the event. The distribution of 
$\vert\cos\theta_T\vert$ is flat for signal events and strongly peaked 
at 1 for background events.  We require $\vert\cos\theta_T\vert<0.9$.

\section{Signal simulation}

In order to take into account final state radiation (FSR) from the 
final state leptons, the matrix element for the decay
$J/\psi\to e^+e^-(\gamma)$ from Ref. \cite{psipaper}, modified
in order to simulate $B^0\to l^+l^-(\gamma)$ with
$l^+l^-=e^+e^-,\mu^+\mu^-$ is included in the event generator.
The detector simulation program is based on GEANT~\cite{geant},
the simulated events are processed as the data.
The overall efficiencies for detecting the decays $B^0\rightarrow e^+e^-$ 
and $B^0\rightarrow \mu^+\mu^-$ are 37.3\% and 46.2\%, respectively, 
without FSR. Including FSR reduces these efficiencies to 31.1\% and 42.4\%, 
respectively.  
We use the efficiencies obtained with FSR 
included for the derivation of limits on branching fractions.

The efficiency for the decay $B^0\rightarrow e^{\pm}\mu^{\mp}$ is 
43.6\% without FSR. There is no matrix element available for the
decay $B^0\rightarrow e^{\pm}\mu^{\mp}(\gamma)$. 
In this mode we use
Monte Carlo without FSR. 
Since final state
radiation reduces the efficiencies for $B^0\rightarrow e^+e^-$ and
$B^0\rightarrow \mu^+\mu^-$ by 20\% and 9\%, respectively,
we estimate the effect of FSR in $B^0\rightarrow e^{\pm}\mu^{\mp}$ 
to be 15\% and include this in the derivation of the upper limit
on the branching fraction as a systematic uncertainty. 

\section{Results}

The background from the process $e^+e^-\to q{\bar q}$ 
results from misidentification of hadrons as leptons and 
has been 
determined as follows.
First, the analysis is redone without imposing the lepton 
identification requirements 
to determine the total number of 2-track events passing all other 
selection criteria 
in the signal and sideband regions 
($5.260  \mbox{ GeV}< M(B)<5.270\mbox{ GeV}$ and 
$\vert \Delta E\vert < 200 \mbox{ MeV}$)
for both the on-resonance 
and off-resonance data samples.  These are used to determine the 
expected number of events in the on-resonance signal region.
(This is done by multiplying the number of events seen in the signal
region in off-resonance data by the ratio of numbers of events 
seen in the sideband region in on- and off-resonance data.)
The probabilities for these tracks to be identified 
as leptons is determined from the relative abundances of particles 
($\pi, K$, proton, electron, muon) and the fake rates. 
The particle abundances for the $B^0$ candidate tracks 
in $q{\bar q}$ events have been determined
using Monte Carlo and were found to be 
$0.58$ for $\pi^{\pm}$,
$0.25$ for $K^{\pm}$, 
$0.15$ for ${p,{\bar p}}$,
and 
$0.004$ for electrons and muons. 
For $\pi^{\pm}$ and $K^{\pm}$ 
the fake rates are determined experimentally from samples of 
$D^{*+}\to D^0\pi^+\to K^-\pi^+\pi^+$  decays.  Proton fake rates 
are taken from Monte Carlo.  The expected number of 
backgrounds events from 
$e^+e^-\to q{\bar q}$ are
$0.01\pm 0.01$, 
$0.14\pm 0.03$ and 
$0.09\pm 0.03$ for the 
$ee, \mu\mu$ and $e\mu$ channels,
respectively.

Backgrounds from $e^+e^-\to\tau^+\tau^-$ events are estimated using a Monte 
Carlo sample analyzed without lepton identification, 
using the procedure described 
above to determine the expected rate of events with the two candidate tracks 
identified as leptons.
The particle abundances for the $B^0$ candidate tracks 
in simulated $\tau^+\tau^-$ events  were found to be 
$0.74$ for $\pi^{\pm}$,
$0.04$ for $K^{\pm}$,
and
$0.11$ for electrons and muons.
The expected numbers of background events are 
$0.10\pm 0.07$ ,
$0.08\pm 0.06$ and 
$0.40\pm 0.20$ events
for the $ee$, $\mu\mu$ and $e\mu$ 
channels, respectively.  
The background from $B{\bar B}$ events is estimated 
from Monte Carlo and
is found to be less
than 0.01 events in all modes.

We find no events in the signal regions for the $e^+e^-$ and $\mu^+\mu^-$ 
modes and two events in the signal region for the $e^{\pm}\mu^{\mp}$
mode. These events are consistent with the expected 
background.  
To conservatively account for the uncertainty in the efficiency determination,
we reduce the efficiency by the estimated systematic error of 
7.6\% in the $e^+e^-$ and $\mu^+\mu^-$ modes and 16.3\% in the $e^{\pm}\mu^{\mp}$
mode when calculating the upper limits. The latter error is dominated by the 
uncertainty from final state radiation.  A summary of the efficiencies, event 
yields and limits is given in Table \ref{summary}.

\begin{table}[tbhp]
\begin{center}
\caption{ Summary of results and branching fraction upper limits. 
$N_{exp}$ is the number of expected background events, 
$N_{obs}$ is the number of events observed, $N_{UL}$ is the 
number of events corresponding to a 90\% CL limit, and  $\epsilon$ is the 
detection efficiency. The first quoted error on $\epsilon$ is
statistical while the second is systematic. }
\label{summary} 
\begin{tabular}{  | l | c | c| c | c | c | c |   }
\hline
mode         &  $N_{exp}$ & $N_{obs}$ &$N_{UL}$& $\epsilon$ & Upper limit  \\
\hline
$B^0\rightarrow e^+e^-$
             &   $0.11\pm 0.07$ & 0  &  2.30  &  31.1$\pm$0.4$\pm$2.4\% & $8.3\times 10^{-7}$  \\
$B^0\rightarrow \mu^+\mu^-$ 
             &   $0.22\pm 0.07$ & 0  &  2.30  &  42.4$\pm$0.5$\pm$3.2\% & $6.1\times 10^{-7}$  \\
$B^0\rightarrow e^{\pm}\mu^{\mp}$
             &  $0.49\pm 0.20$  & 2  &  5.24  &  43.6$\pm$0.5$\pm$7.1\% & $14.9\times 10^{-7}$  \\
\hline
\end{tabular}
\end{center}
\end{table}


\section{Limits on Pati-Salam leptoquarks}

One of the simplest models that assumes symmetry between
quarks and leptons is the Pati-Salam model\cite{Pati-Salam}.
This model predicts heavy spin-one gauge bosons,
called leptoquarks (LQ), that carry both color
and lepton quantum numbers. While most leptoquark
models assume that the leptoquarks couple within one generation
only,  Pati-Salam model allows 
cross-generation couplings. Consequently,
Pati-Salam leptoquarks can mediate the decay 
$B^0\rightarrow e^{\pm}\mu^{\mp}$ \cite{bemLQ,LQmass}.
The relationship between ${\cal B}(B^0\rightarrow e^{\pm}\mu^{\mp})$
and leptoquark mass is \cite{LQmass}
$$\Gamma ( B^0\rightarrow e^{\pm}\mu^{\mp})=\pi\alpha_s^2(M_{LQ}) 
{1\over M_{LQ}^4} F^2_{B^0} m^3_{B^0} R^2$$
where 
$$R={m_{B^0}\over m_b}
	{\Big( {\alpha_s(M_{LQ})\over \alpha_s(m_t) }\Big)}^{-{4\over 7}}
	{\Big( {\alpha_s(m_t)\over \alpha_s(m_b) }\Big)}^{-{12\over 23}}.
$$
	
The first attempt to constrain the Pati-Salam leptoquark masses
by measuring the branching fraction for $B^0\rightarrow e^{\pm}\mu^{\mp}$
was made by the CDF collaboration\cite{CDFem}. The result
obtained was $M_{LQ}>21.7\mbox{ TeV}$. We follow  the CDF procedure
by using $F_B=175\pm 30$ MeV for $B^0$ decay constant~\cite{FB},
$m_{B^0}=5279.1\pm 0.7\mbox{(stat.)}\pm 0.3\mbox{(syst.)}$ MeV\cite{CLEO_B_mass}, 
$m_t=176.0\pm 6.5$ GeV~\cite{CDF-top-mass}
and obtain  a new limit on the Pati-Salam
leptoquark mass of $M_{LQ}>27\mbox{ TeV}$ at 90\% confidence level.

In conclusion we find no evidence for the 
decays $B^0\to e^+e^-,\mu^+\mu^-$ and 
$e^{\pm}\mu^{\mp}$ and obtain upper limits
on the corresponding 
branching fractions of 
${\cal B}(B^0\rightarrow e^+e^-)< 8.3\times 10^{-7}$, 
${\cal B}(B^0\rightarrow  \mu^+\mu^-)< 6.1\times 10^{-7}$,
${\cal B}(B^0\rightarrow e^{\pm}\mu^{\mp})< 15\times 10^{-7}$ at 
90\% confidence level. The limit on 
${\cal B}(B^0\rightarrow e^{\pm}\mu^{\mp})$
corresponds to a lower limit of 27 TeV on the Pati-Salam leptoquark mass
at 90\% CL.


\begin{figure}
  \begin{center}
   \mbox{\epsfig{file=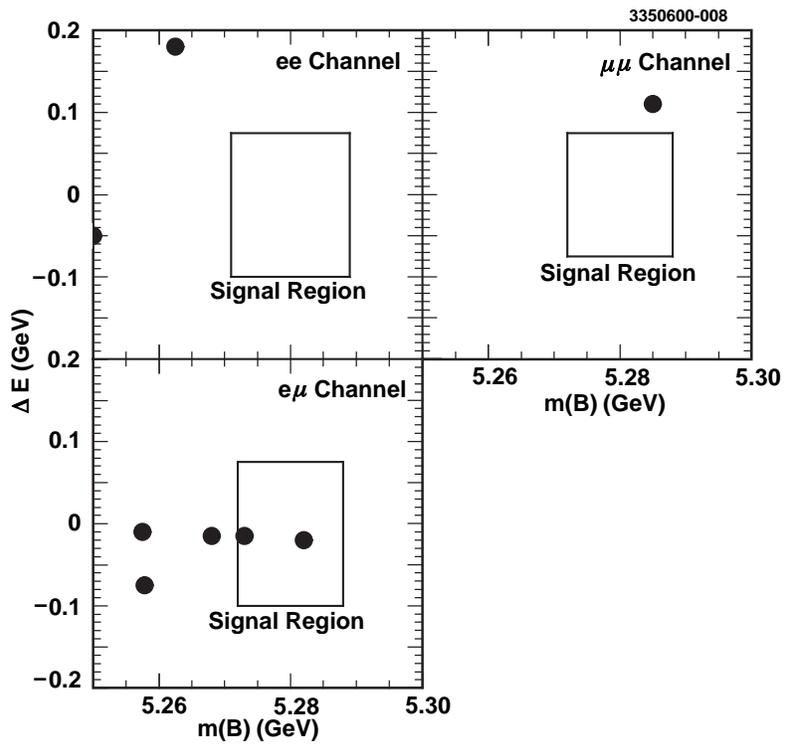,height=10cm}}
\caption{
\label{mde} Beam constrained mass of dilepton pair
vs energy difference 
$\Delta E = E_{l^+}+E_{l^-}-E_{beam}$  for
on resonance data. All selection criteria 
except the requirements on
lepton pair mass and $\Delta E$ 
are included. The box in the center
of each plot shows the signal region.
}
  \end{center}
\end{figure}


We gratefully acknowledge the effort of the CESR staff in providing us with
excellent luminosity and running conditions.
I.P.J. Shipsey thanks the NYI program of the NSF, 
M. Selen thanks the PFF program of the NSF, 
M. Selen and H. Yamamoto thank the OJI program of DOE, 
M. Selen and V. Sharma 
thank the A.P. Sloan Foundation, 
M. Selen and V. Sharma thank the Research Corporation, 
F. Blanc thanks the Swiss National Science Foundation, 
and H. Schwarthoff and E. von Toerne
thank the Alexander von Humboldt Stiftung for support.  
This work was supported by the National Science Foundation, the
U.S. Department of Energy, and the Natural Sciences and Engineering Research 
Council of Canada.



\begin{thebibliography}{99}








\bibitem{SM-br} 

A. Ali, C. Greub and T. Mannel, in 
{\it Proceedings of the ECFA Workshop on the 
Physics of the European B Meson factory },
Edited by
R. Aleksan and A. Ali. European Committee for Future
Accelerators, 1993 (155).  (ECFA-93-151,C93/03/26)





\bibitem{Rizzo} 

See, for example,
%
J. L. Hewett, S. Nandi and T. G. Rizzo, 
Phys. Rev. D {\bf 39}, 250 (1989);
%
X.G. He, T.D. Nguyen and R.R. Volkas, Phys. Rev. D {\bf 38}, 814 (1988);
%
W. Skiba and J. Kalinowski, Nucl. Phys. B {\bf 404}, 3 (1993);
%
Y.-B. Dai, C.-S. Huang and W.-W. Huang, Phys. Lett. B {\bf 390}, 257 (1997);
%
H.E. Logan and U. Nierste, hep-ph/0004139,FERMILAB-Pub-00/084-T.

\bibitem{Nir} 

Y. Nir and D. Silverman, Phys. Rev. D {\bf 42}, 1447 (1990).

\bibitem{Gronau} 

M. Gronau and D. London, Phys. Rev. D {\bf 55}, 2845 (1997).


\bibitem{Pati-Salam} 

J. Pati and A. Salam, Phys. Rev. D {\bf 10}, 275 (1974).

\bibitem{lingel-paper} 

R. Ammar {\it et al}. (CLEO collaboration) Phys. Rev. D {\bf 49}, 5701 (1994).

\bibitem{CDFem} 
F. Abe {\it et al}. (CDF collaboration) Phys. Rev. Lett. {\bf 81}, 5742 (1998). 



\bibitem{CDFmm} 

F. Abe {\it et al,} (CDF collaboration), Phys. Rev. D {\bf 57}, R3811 (1998).

\bibitem{cleo_detector} 

Y.Kubota {\it et al},(CLEO collaboration) Nucl. Instr. Methods Phys. Res., 
Sect A {\bf 320}, 66 (1992);
%
T.Hill, Nucl. Instr. Methods Phys. Res., 
Sect A {\bf 418}, 32 (1998). 

\bibitem{psipaper} 
T. Armstrong {\it et al},(E760 collaboration) Phys. Rev D {\bf 54}, 7067 (1996).

\bibitem{geant} CERN Program Library Long Writeup W5013 (1993).

\bibitem{bemLQ} 
A.V. Kuznetsov and N.V. Mikheev, Phys. Lett. B {\bf 329}, 295 (1994).

\bibitem{LQmass} 
G. Valencia and S. Willenbrock, Phys. Rev. D {\bf 50}, 6843 (1994).



\bibitem{FB}C. Bernard, Proceedings of 7th International Symposium on
Heavy Flavor Physics, Santa Barbara CA, 7-11 July 1997, 
Edited by C. Campagnari. Singapore, World Scientific, 1999.


\bibitem{CDF-top-mass}
F. Abe {\it et al},(CDF collaboration) Phys. Rev. Lett. {\bf 82}, 271 (1999). 

\bibitem{CLEO_B_mass}
 S. E. Csorna {\it et al}, (CLEO Collaboration) Phys.Rev. D {\bf 61} 111101(2000).











\end{thebibliography}
\end{document}